\RequirePackage{lineno}

\documentclass[aps,prc,twocolumn,groupedaddress,showpacs,amsmath,amssymb,floatfix,superscriptaddress]{revtex4}
\usepackage{multirow}
\usepackage{graphicx}
\usepackage{times}
\begin{document}

%\linenumbers

%%%%%%%%%%%%%%%%%%%%%%%%%%%%%
%  PLEASE SET THE DEFAULT FIGURE ANGLE HERE
%%%%%%%%%%%%%%%%%%%%%%%%%%%%%

\newcommand{\DefaultFigureAngle}{0}
%\newcommand{\DefaultFigureAngle}{270}

%%%%%%%%%%%%%%%%%%%%%%%%%%%%%
%%%%%%%%%%%%%%%%%%%%%%%%%%%%%

\title{Constraints on $\theta_{13}$ from A Three-Flavor Oscillation Analysis of \\
Reactor Antineutrinos at KamLAND}

% All university affiliations addresses go here:
\newcommand{\tohoku}{\affiliation{Research Center for Neutrino
    Science, Tohoku University, Sendai 980-8578, Japan}}
\newcommand{\alabama}{\affiliation{Department of Physics and
    Astronomy, University of Alabama, Tuscaloosa, Alabama 35487, USA}}
\newcommand{\lbl}{\affiliation{Physics Department, University of
    California, Berkeley, and \\ Lawrence Berkeley National Laboratory, 
Berkeley, California 94720, USA}}
\newcommand{\caltech}{\affiliation{W.~K.~Kellogg Radiation Laboratory,
    California Institute of Technology, Pasadena, California 91125, USA}}
\newcommand{\colostate}{\affiliation{Department of Physics, Colorado
    State University, Fort Collins, Colorado 80523, USA}}
\newcommand{\drexel}{\affiliation{Physics Department, Drexel
    University, Philadelphia, Pennsylvania 19104, USA}}
\newcommand{\hawaii}{\affiliation{Department of Physics and Astronomy,
    University of Hawaii at Manoa, Honolulu, Hawaii 96822, USA}}
\newcommand{\kansas}{\affiliation{Department of Physics,
    Kansas State University, Manhattan, Kansas 66506, USA}}
\newcommand{\lsu}{\affiliation{Department of Physics and Astronomy,
    Louisiana State University, Baton Rouge, Louisiana 70803, USA}}
\newcommand{\stanford}{\affiliation{Physics Department, Stanford
    University, Stanford, California 94305, USA}}
\newcommand{\ut}{\affiliation{Department of Physics and
    Astronomy, University of Tennessee, Knoxville, Tennessee 37996, USA}}
\newcommand{\tunl}{\affiliation{Triangle Universities Nuclear Laboratory and Physics Departments, Duke University, Durham, North Carolina 27708, USA; \\ 
North Carolina Central University, Durham, North Carolina 27707, USA; \\
and the University of North Carolina at Chapel Hill, Chapel Hill, North Carolina 27599, USA}}
\newcommand{\wisc}{\affiliation{Department of Physics, University
    of Wisconsin, Madison, Wisconsin 53706, USA}}  
\newcommand{\cnrs}{\affiliation{CEN Bordeaux-Gradignan, IN2P3-CNRS and
    University Bordeaux I, F-33175 Gradignan Cedex, France}}
\newcommand{\ipmu}{\affiliation{Institute for the Physics and Mathematics of the 
    Universe, Tokyo University, Kashiwa 277-8568, Japan}}
\newcommand{\nikhef}{\affiliation{Nikhef, Science Park 105, 1098 XG Amsterdam, the Netherlands}}

% Put Present addresses here:
    
\newcommand{\atlanlnow}{\altaffiliation{Present address: Subatomic Physics Group, Los Alamos National Laboratory, Los Alamos, NM 87545, USA}}
    
\newcommand{\atksunow}{\altaffiliation{Present address: Department of Physics, Kansas State University, Manhattan, Kansas 66506, USA}}

\newcommand{\atokayamanow}{\altaffiliation{Present address: Department of Physics, Tokyo Institute of Technology, Tokyo 152-8551, Japan}}

\newcommand{\atregisnow}{\altaffiliation{Present address: Department of Physics and Computational Science, Regis University, Denver, Colorado 80221, USA}}

\newcommand{\atfnalnow}{\altaffiliation{Present address: Fermi National Accelerator Laboratory, Batavia, Illinois 60510, USA}}

\newcommand{\atsnolabnow}{\altaffiliation{Present address: SNOLAB, Lively, ON P3Y 1N2, Canada}}

\newcommand{\atllnlnow}{\altaffiliation{Present address: Lawrence Livermore National Laboratory, Livermore, California 94550, USA}}

\newcommand{\atucdnow}{\altaffiliation{Present address: Department of Physics, University of California, Davis, California 95616, USA}}

\newcommand{\atuwnow}{\altaffiliation{Present address:  CENPA, University of Washington, Seattle, Washington 98195, USA}}

\newcommand{\atumdnow}{\altaffiliation{Present address: Department of Physics, University of Maryland, College Park, Maryland 20742, USA}}

\newcommand{\atmitnow}{\altaffiliation{Present address: Department of Physics, Massachusetts Institute of Technology, Cambridge, MA 02139, USA}}

\newcommand{\atrowannow}{\altaffiliation{Present address: Department of Physics and Astronomy, Rowan University, 201 Mullica Hill Road, Glassboro, New Jersey 08028, USA}}

\newcommand{\atsdnow}{\altaffiliation{Present address: Department of Physics, University of South Dakota, 414 E. Clark St. Vermillion, South Dakota 57069, USA}}

\newcommand{\atuwjnow}{\altaffiliation{Jointly at: Center for Experimental Nuclear Physics and Astrophysics, University of Washington, Seattle, Washington 98195, USA}}

\newcommand{\OscPhaseEff}{\Delta m^{2}\mathbb{L}/\mathbb{E})_{\rm{Eff}}}

%
% Note: some authors have joint appointments with IPMU (http://www.ipmu.jp/members/)
%
% Tohoku
\author{A.~Gando}\tohoku
\author{Y.~Gando}\tohoku
\author{K.~Ichimura}\tohoku
\author{H.~Ikeda}\tohoku
\author{K.~Inoue}\tohoku\ipmu
\author{Y.~Kibe}\atokayamanow\tohoku
\author{Y.~Kishimoto}\tohoku
\author{M.~Koga}\tohoku\ipmu
\author{Y.~Minekawa}\tohoku
\author{T.~Mitsui}\tohoku
\author{T.~Morikawa}\tohoku
\author{N.~Nagai}\tohoku
\author{K.~Nakajima}\tohoku
\author{K.~Nakamura}\tohoku\ipmu
\author{K.~Narita}\tohoku
\author{I.~Shimizu}\tohoku
\author{Y.~Shimizu}\tohoku
\author{J.~Shirai}\tohoku
\author{F.~Suekane}\tohoku
\author{A.~Suzuki}\tohoku
\author{H.~Takahashi}\tohoku
\author{N.~Takahashi}\tohoku
\author{Y.~Takemoto}\tohoku
\author{K.~Tamae}\tohoku
\author{H.~Watanabe}\tohoku
\author{B.D.~Xu}\tohoku
\author{H.~Yabumoto}\tohoku
\author{H.~Yoshida}\tohoku
\author{S.~Yoshida}\tohoku
% 
% IPMU
\author{S.~Enomoto}\atuwjnow\ipmu % tohoku --> ipmu
\author{A.~Kozlov}\ipmu
\author{H.~Murayama}\ipmu\lbl
%
% Alabama
\author{C.~Grant}\alabama
\author{G.~Keefer}\atllnlnow\alabama
\author{A.~Piepke}\ipmu\alabama
%
% LBL and UC Berkeley
\author{T.I.~Banks}\lbl
\author{T.~Bloxham}\lbl
\author{J.A.~Detwiler}\lbl
\author{S.J.~Freedman}\ipmu\lbl
\author{B.K.~Fujikawa}\ipmu\lbl
\author{K.~Han}\lbl
\author{R.~Kadel}\lbl
\author{T.~O'Donnell}\lbl
\author{H.M.~Steiner}\lbl
%
% Caltech
\author{D.A.~Dwyer}\caltech
\author{R.D.~McKeown}\caltech
\author{C.~Zhang}\caltech
%
% Colorado State
\author{B.E.~Berger}\colostate
%
% Drexel
\author{C.E.~Lane}\drexel
\author{J.~Maricic}\drexel
\author{T.~Miletic}\atrowannow\drexel
%
% Hawaii
\author{M.~Batygov}\atsnolabnow\hawaii
\author{J.G.~Learned}\hawaii
\author{S.~Matsuno}\hawaii
\author{M.~Sakai}\hawaii
%
% KSU
\author{G.A.~Horton-Smith}\ipmu\kansas
%
% Stanford
\author{K.E.~Downum}\stanford
\author{G.~Gratta}\stanford
%
% UT
\author{Y.~Efremenko}\ipmu\ut
\author{O.~Perevozchikov}\atsdnow\ut
%
% TUNL
\author{H.J.~Karwowski}\tunl
\author{D.M.~Markoff}\tunl
\author{W.~Tornow}\tunl
%
% Wisconsin
\author{K.M.~Heeger}\ipmu\wisc
% 
% NIKHEF
\author{M.P.~Decowski}\ipmu\lbl\nikhef

\collaboration{The KamLAND Collaboration}\noaffiliation

\date{\today}

\begin{abstract}

We present new constraints on the neutrino oscillation parameters $\Delta m^{2}_{21}$, $\theta_{12}$, and $\theta_{13}$ from a three-flavor analysis of solar and KamLAND data. The KamLAND data set includes data acquired following a radiopurity upgrade and amounts to a total exposure of  $3.49 \times 10^{32}$ target-proton-year.  Under the assumption of {\it CPT} invariance, a two-flavor analysis (\mbox{$\theta_{13} = 0$}) of the KamLAND and solar data yields the best-fit values $\tan^{2} \theta_{12} = 0.444^{+0.036}_{-0.030}$ and $\Delta m^{2}_{21} = 7.50^{+0.19}_{-0.20} \times 10^{-5} ~ {\rm eV}^{2}$; a three-flavor analysis with $\theta_{13}$ as a free parameter yields the best-fit values $\tan^{2} \theta_{12} = 0.452^{+0.035}_{-0.033}$, $\Delta m^{2}_{21} = 7.50^{+0.19}_{-0.20} \times 10^{-5} ~ {\rm eV}^{2}$, and $\sin^{2} \theta_{13} = 0.020^{+0.016}_{-0.016}$. This $\theta_{13}$ interval is consistent with other recent work combining the CHOOZ, atmospheric, and long-baseline accelerator experiments.  We also present a new global $\theta_{13}$ analysis, incorporating the CHOOZ, atmospheric and accelerator data, which indicates $\sin^{2} \theta_{13} = 0.009^{+0.013}_{-0.007}$.
A nonzero value is suggested, but only at the 79\% C.L.

\end{abstract}

\pacs{14.60.Pq, 26.65.+t, 28.50.Hw, 91.35.-x}

\maketitle

\section{Introduction}
\label{section:Introduction}
Neutrino flavor oscillation is by now well established by the convergence of results from experiments involving solar, reactor, atmospheric, and accelerator neutrinos. Central to any discussion of neutrino oscillation phenomenology is the Pontecorvo-Maki-Nakagawa-Sakata (PMNS) mixing matrix which describes neutrino mixing in analogy to the Cabibbo-Kobayashi-Maskawa (CKM) matrix of the quark sector~\cite{Nakamura2010}. Although the possibility of more than three neutrino mass states, motivated in part by~\cite{Aguilar2001}, is not excluded, our notation and discussion is restricted to the assumption of three neutrino mass states. In this case, the three-flavor eigenstates ($\nu_{e}\,,\nu_{\mu}\,,\nu_{\tau}$) can be expressed as a linear combination of the three mass eigenstates ($\nu_{1}\,,\nu_{2}\,,\nu_{3}$)\,:
\vspace{-0.47cm}
 \[
\left| \nu_{\alpha}\right\rangle = \sum_{i=1}^{3} U_{\alpha i} \left| \nu_{i}\right\rangle \hspace{1cm} (\alpha = e, \mu, \tau)\;. \label{Eq-FlavorMass}
\]
Ignoring possible Majorana phases which are irrelevant to oscillation phenomenology, the PMNS matrix $U$ is parametrized by three mixing angles, $\theta_{12}$, $\theta_{23}$, $\theta_{13}$, and a {\it CP}-violating phase $\delta$. $U$ may be written as 
\begin{eqnarray}
\label{eqn:MSNPMatrix}
U & = & 
\left(
\begin{array}{ccc}
1 & 0 & 0\\
0 & c_{23} & s_{23}\\
0 & -s_{23} & c_{23}\\
\end{array}
\right)
\left(
\begin{array}{ccc}
c_{13} & 0 & s_{13}e^{-i\delta}\\
0 & 1 & 0\\
-s_{13}e^{i\delta} & 0 & c_{13}\\
\end{array}
\right) \nonumber \\
 & & 
\times
\left(
\begin{array}{ccc}
c_{12} & s_{12} & 0\\
-s_{12} & c_{12} & 0\\
0 & 0 & 1\\
\end{array}
\right)\;,
\end{eqnarray}
where $s_{ij} = \sin \theta_{ij}$ and $c_{ij} = \cos \theta_{ij}$.

The mass-squared splittings ($\Delta m^{2}_{ij} \equiv m^2_i-m^2_j$) between the neutrino mass states are described by two independent parameters, $\Delta m^{2}_{21}$ and $\Delta m^{2}_{32}$. At the currently achieved sensitivity, mixing between $\nu_{1}$ and $\nu_{2}$ ($\nu_{1}$-$\nu_{2}$ mixing) can explain the KamLAND data~\cite{Abe2008} and also, with addition of MSW enhancement~\cite{Wolfenstein1979, Mikheyev1985}, the solar results~\cite{Cleveland1998, Hampel1999, Altmann2005, Abdurashitov2009, Hosaka2006, Aharmim2010, Aharmim2008, Arpesella2008b}.  Atmospheric~\cite{Wendell2010}, K2K~\cite{Ahn2006}, and MINOS~\cite{Adamson2008} data can be accommodated by $\nu_{2}$-$\nu_{3}$ mixing.  As of yet, there is no experimental evidence of $\nu_{1}$-$\nu_{3}$ mixing (i.e., a nonzero $\theta_{13}$) with high statistical significance.

Probing the value of $\theta_{13}$ is a subject of intense ongoing activity. The most stringent limit to date, from the 1-km-baseline CHOOZ reactor experiment~\cite{Apollonio2003}, is $\sin^{2} \theta_{13} < 0.04$ at the 90\% C.L. Next-generation accelerator experiments (T2K~\cite{Itow2001} and NO$\nu$A~\cite{Ayres2005}) and reactor experiments (Double Chooz~\cite{Ardellier2004}, Daya Bay~\cite{Guo2007}, and RENO~\cite{Ahn2010}) aim to significantly improve the sensitivity to this parameter and may definitively determine the value of $\theta_{13}$.  If $\theta_{13}$ is nonzero, future oscillation experiments may explore leptonic {\it CP} violation (parametrized by $\delta$) and probe the neutrino mass hierarchy (i.e., the sign of $\Delta m^{2}_{32}$). The feasibility of such experiments and the path forward depend critically on the magnitude of $\theta_{13}$. 
  
This article presents an updated KamLAND data set and focuses on new constraints on $\theta_{12}$, $\Delta m^{2}_{21}$, and $\theta_{13}$ based on a three-flavor combined analysis of KamLAND and solar data. As motivated by~\cite{Fogli2008}, we also present a global analysis including the CHOOZ, accelerator, and atmospheric oscillation experiments in order to explore possible hints of nonzero $\theta_{13}$.  

\section{Approximate Three-flavor neutrino oscillation formalism }
\label{section:Earth}
Previous KamLAND results~\cite{Abe2008} were based on a two-flavor ($\nu_{1}$-$\nu_{2}$) oscillation formalism which assumes \mbox{$\theta_{13}=0$}.  For the length scales relevant to reactor neutrino oscillation at KamLAND and solar neutrino oscillation in the LMA-MSW solution,  the dependence of the more general three-flavor phenomenology on the larger $\nu_{1}$-$\nu_{3}$ mass splitting \mbox{$(|\Delta m^2_{31}| \sim |\Delta m^2_{32}|\gg \Delta m^2_{21})$} averages out and the three-flavor survival probability ($P^{3\nu}_{ee}$), including matter effects, may be approximated as 
\begin{eqnarray}
P_{ee}^{3\nu} = \cos^{4}\theta_{13} \widetilde{P}_{ee}^{2\nu} + \sin^{4}\theta_{13} \;.
\end{eqnarray}
$\widetilde{P}_{ee}^{2\nu}$ has the same form as the survival probability in matter for \mbox{$\nu_{1}$-$\nu_{2}$} mixing but with the electron density \mbox{($N_{e}$)} modified: \mbox{$\widetilde{N}_{e} = N_{e}\cos^{2}\theta_{13}$}~\cite{Goswami2005}. Since \mbox{$\sin^{2}\theta_{13} \ll 1$}, the survival probability can be further approximated as \mbox{$P_{ee}^{3\nu} \sim (1 - 2 \sin^{2}\theta_{13}) \widetilde{P}_{ee}^{2\nu}$}.  Thus, for KamLAND and the solar experiments, $\nu_{1}$-$\nu_{3}$ mixing would give rise to an energy-independent suppression of the survival probability relative to the $\theta_{13}=0$ case.

For solar neutrino oscillation in the LMA-MSW solution, coherent mixing can be safely ignored due to the long distance between the Sun and the Earth. The two-neutrino survival probability is simply expressed as
\begin{eqnarray}
\widetilde{P}_{ee}^{2\nu} = P^{\odot}_{1} P_{1e} + P^{\odot}_{2} P_{2e} \;,
\end{eqnarray}
where $P^{\odot}_{i}$ and $P_{ie}$ are, respectively, the probability of the \mbox{$\nu_{e} \rightarrow \nu_{i}$} transition in the Sun and the probability of the \mbox{$\nu_{i} \rightarrow \nu_{e}$} transition in the Earth with the modified electron density \mbox{$\widetilde{N}_{e}$}.  Neutrino propagation in the Sun and Earth is calculated following the analytical procedure of~\cite{Parke1986, Lisi1997}, and the resulting survival probabilities agree well with numerical calculations.

For reactor antineutrinos studied at KamLAND, the matter effect in the Earth is not as large as for solar neutrinos. Assuming a constant rock density (2.7 $\rm{g}/\rm{cm}^{3}$), the two-neutrino survival probability is given by
\begin{eqnarray}
\widetilde{P}_{ee}^{2\nu} = 1 - \sin^{2} 2\theta_{12M} \sin^{2} \left( \frac{1.27 \Delta m_{21M}^{2} L}{E} \right)\,, \label{Eq-P-2nu-KamLAND}
\end{eqnarray}
where 
$L$ is the electron antineutrino ($\overline{\nu}_{e}$) flight distance in meters from the source to the detector, $E$ is the $\overline{\nu}_{e}$ energy in MeV, and $\Delta m_{21}^{2}$ is in ${\rm eV}^{2}$.  $\theta_{12M}$ and $\Delta m^2_{21M}$ are the matter-modified mixing angle and mass splitting defined by
\begin{eqnarray}
\label{Eq-Theta12M}
\sin^{2} 2\theta_{12M} = \frac{\sin^{2} 2\theta_{12}}{(\cos 2\theta_{12} - A / \Delta m_{21}^{2})^{2} + \sin^{2} 2\theta_{12}}\;,\\ 
\Delta m_{21M}^{2} = \Delta m_{21}^{2} \sqrt{(\cos 2\theta_{12} - A / \Delta m_{21}^{2})^{2} + \sin^{2} 2\theta_{12}}\;.
\end{eqnarray}
$A = -2 \sqrt{2} G_{F} \widetilde{N}_{e} E$, and has a negative sign for antineutrinos; $G_{F}$ is the Fermi constant. The matter effect modifies the expected reactor $\overline{\nu}_{e}$ event rate by up to 3\%, depending on the oscillation parameters.

\section{KamLAND experiment}
\label{section:Detector}
The KamLAND detector is located in Kamioka mine, Gifu, Japan. The primary target volume consists of 1~kton of ultra-pure liquid scintillator~(LS).  This inner detector~(ID) of LS is shielded by a 3.2-kton water-Cherenkov outer detector (OD).  Scintillation light is viewed by 1325 17-inch and 554 20-inch photomultiplier tubes (PMTs) providing 34\% solid-angle coverage.  A detailed overview of the detector is given in~\cite{Abe2010}.  

The $\overline{\nu}_{e}$ flux at KamLAND is dominated by 56 Japanese nuclear power reactors. The flux-weighted average baseline to these reactors is $\sim$180 km.  The reactor fluxes are calculated precisely based on detailed operational data including the thermal power variation and fuel replacement and reshuffling records, provided for all Japanese commercial reactors by a consortium of Japanese electric power companies. The absolute thermal power, used to normalize the fission rates, is measured to within 2\% for each reactor. This uncertainty is conservatively assumed to be correlated across all reactors, though some potentially uncorrelated components have been put forward in~\cite{Djurcic2009}. The data points are typically provided at weekly frequency during regular operations when the relative instability is of the order of $10^{-3}$. When the operating parameters vary more quickly, the data are provided at higher frequency, with a period between 10 min and 1 h. The relative fission yields, averaged over the entire live-time period, for isotopes ($^{235}$U : $^{238}$U : $^{239}$Pu : $^{241}$Pu) are (0.571 : 0.078 : 0.295 : 0.056), respectively. The detailed reactor operation data are also used for accurate tracking of the flux-weighted average reactor baseline and spectrum shape change over the course of the experiment.  The contribution from Korean reactors, based on reported electric power generation, is estimated to be $(3.4 \pm 0.3)\%$ . The contribution from Japanese research reactors and the remainder of the global nuclear power industry, estimated using reactor specifications from the International Nuclear Safety Center~\cite{INSC2010}, is $(1.0 \pm 0.5)\%$. The $\overline{\nu}_{e}$ spectra per fission provided in~\cite{Schreckenbach1985, Hahn1989, Vogel1981} are used, and the uncertainties are further constrained from~\cite{Achkar1996}. In addition, the long-lived, out-of-equilibrium fission products $^{90}$Sr, $^{106}$Ru, and $^{144}$Ce~\cite{Kopeikin2001} are evaluated from the history of fission rates for each isotope and are found to contribute only $(0.6 \pm 0.3)\%$.

Electron antineutrinos are detected in KamLAND via the inverse beta-decay reaction, \mbox{$\overline{\nu}_{e}+p\rightarrow e^{+}+n$}. This process has a delayed coincidence~(DC) event pair signature which offers powerful background suppression.  The energy deposited by the positron generates the DC pair's prompt event and is approximately related to the incident $\overline{\nu}_{e}$ energy by \mbox{$E \simeq E_{\rm p} + \overline{E}_{n} + 0.8 ~{\rm MeV}$, where $E_{\rm p}$} is the sum of the $e^{+}$ kinetic energy and annihilation $\gamma$ energies, and $\overline{E}_{n}$ is the average neutron recoil energy, $O(10 ~\rm{keV})$. The delayed event in the DC pair is generated by the capture $\gamma$ produced when the neutron captures on a proton or $^{12}{\rm C}$ nucleus. The mean neutron capture time is $207.5 \pm 2.8 ~\mu \rm{s}$~\cite{Abe2010}\,.

In the previous KamLAND result~\cite{Abe2008} the largest background in the prompt energy region below 3.0 MeV came from $^{13}{\rm C}(\alpha,{\it n})^{16}{\rm O}$ reactions induced by $\alpha$-decays in the LS.  This affected the estimation of the flux of geologically produced antineutrinos (geo-$\overline{\nu}_{e}$) expected between 0.9 MeV and 2.6 MeV from the decay chains of $^{238}$U and $^{232}$Th in the Earth~\cite{Araki2005b, Bellini2010}.  In 2007 the KamLAND collaboration started a campaign to purify the LS and ultimately achieved a twenty-fold reduction of $^{210}$Po, the dominant $\alpha$-decay source.  This reduction gives a better signal-to-background ratio for the geo-$\overline{\nu}_{e}$ flux estimation and enhances sensitivity to reactor $\overline{\nu}_{e}$ oscillations below 2.6 MeV.  

We present an improved measurement of reactor $\overline{\nu}_{e}$ oscillation based on data collected from March 9, 2002, to November 4, 2009.  This sample includes the previously reported data set~\cite{Abe2008}, denoted hereafter as \mbox{DS-1}, in addition to data collected after LS purification commenced, designated as \mbox{DS-2}.  The total live time is 2135 days after removing periods of low data quality which occurred during LS purification, and after detector vetoes to reduce cosmogenic backgrounds.  The high-quality data selected from \mbox{DS-2} accounts for 30.41\% of the total live time.  The number of target protons within the 6.0-m-radius spherical fiducial volume is calculated to be $(5.98 \pm 0.12) \times 10^{31}$ for the combined data set, which corresponds to an exposure to $\overline{\nu}_{e}$ of $3.49 \times 10^{32}$ proton-years. 

Physical quantities such as event vertex and energy are reconstructed based on the timing and charge distributions of scintillation photons recorded by the ID PMTs.  The vertex and energy reconstructions are calibrated using $^{60}$Co, $^{68}$Ge, $^{203}$Hg, $^{65}$Zn, $^{241}$Am$^{9}$Be, $^{137}$Cs, and $^{210}$Po$^{13}$C radioactive sources. The observed vertex resolution is $\sim$12 ${\rm cm}/\sqrt{E{\rm (MeV)}}$, and the energy resolution is $6.4\%/\sqrt{E{\rm (MeV)}}$. For \mbox{DS-2}, the resolutions are time-dependent due to a light-yield reduction of up to $\sim$20\% relative to \mbox{DS-1}. 
The source calibrations are augmented with studies of muon spallation products to monitor the detector stability and to determine the nonlinearity of the energy response due to LS quenching, Cherenkov light, and dark hit contributions.
The systematic uncertainty of the absolute energy response over the full \mbox{DS-1} and \mbox{DS-2} data sets is less than 1.2\%, and when propagated in the reactor $\overline{\nu}_{e}$ spectrum produces a 1.8\% uncertainty on $\Delta m^{2}_{21}$ and a 1.3\% uncertainty on the event rate above the analysis threshold.  

For \mbox{DS-1}, the systematic uncertainty on the fiducial volume up to 5.5 m radius was determined to be 1.6\% with a full-volume calibration campaign~\cite{Bruce2009}.  The uncertainty in the volume between 5.5 m and 6.0 m radius was estimated from the vertex uniformity of muon-induced $^{12}$B and $^{12}$N; the combined uncertainty on the 6.0-m-radius fiducial volume for \mbox{DS-1} is 1.8\%.  To date there have been no full-volume calibrations for \mbox{DS-2}, so we rely on vertex uniformity of cosmogenic $^{12}$B and $^{12}$N events; in this case, we assign a 2.5\% uncertainty on the 6.0-m-radius fiducial volume. 

Table~\ref{table:systematic} summarizes the systematic uncertainties on $\Delta m^{2}_{21}$ and the expected event rate of reactor $\overline{\nu}_{e}$'s; the overall rate uncertainties for \mbox{DS-1} and \mbox{DS-2} are 4.1\% and 4.5\%, respectively.

\begin{center}
\begin{table}[t]
\caption{\label{table:systematic}Estimated systematic uncertainties for the neutrino oscillation parameters $\Delta m^{2}_{21}$, $\theta_{12}$, and $\theta_{13}$ for the earlier/later periods of measurement, denoted in the text as \mbox{DS-1/DS-2}. The overall uncertainties are 4.1\% / 4.5\% for \mbox{DS-1/DS-2}.
}

 \begin{tabular}{@{}*{7}{l}}
 \hline
 \hline
  & Detector-related (\%) \hspace{-0.5cm} &  & Reactor-related (\%) \hspace{-0.5cm} & \\
 \hline
 $\Delta m^{2}_{21}$ \hspace{0.0cm} & Energy scale &  1.8 / 1.8  &  $\overline{\nu}_{e}$-spectra~\cite{Achkar1996} & \hspace{0.2cm} 0.6 / 0.6 \vspace{0.1cm}\\
 Rate \hspace{0.2cm} & Fiducial volume & 1.8 / 2.5 &  $\overline{\nu}_{e}$-spectra & \hspace{0.2cm} 2.4 / 2.4\\
  & Energy scale & 1.1 / 1.3 & Reactor power & \hspace{0.2cm} 2.1 / 2.1 \\
  & $L_{cut}(E_{\rm p})$ eff. & 0.7 / 0.8 & Fuel composition & \hspace{0.2cm} 1.0 / 1.0\\
  & Cross section & 0.2 / 0.2 & Long-lived nuclei & \hspace{0.2cm} 0.3 / 0.4\\
% \hline
 & Total \hspace{0.0cm} & 2.3 / 3.0 \hspace{0.2cm} & Total &\hspace{0.2cm} 3.3 / 3.4\\
 \hline
 \hline
 \end{tabular}
 \end{table}
 \end{center}

\vspace{-0.8cm}

\section{ KamLAND data reduction and candidate event selection }
\label{section:cutsAndBkg}
\begin{center}
\begin{table}[b]
\caption{\label{table:background}Estimated backgrounds excluding geo-$\overline{\nu}_{e}$ after first- and second-level cuts.
}
\centering
\begin{tabular}{@{}*{7}{llr@{}l@{}c@{}r@{}l}}
\hline
\hline
Background \hspace{-2.2cm} & & \multicolumn{5}{c}{Contribution} \\
\hline
1 & Accidental & 102.&5 & ~$\pm$~ & 0.&1\\
2 & $^{9}$Li/$^{8}$He & 24.&8 & $\pm$ & 1.&6\\
\multirow{2}{*}{3 $\bigg{\lbrace}$}  
& $^{13}{\rm C}(\alpha,{\it n})^{16}{\rm O}_{\rm g.s.}$, $np \rightarrow np$ & 171.&7 & $\pm$ & 18.&2\\
& $^{13}{\rm C}(\alpha,{\it n})^{16}{\rm O}_{\rm g.s.}$, $^{12}{\rm C}({\it n},{\it n'})^{12}{\rm C}^{*}$ (4.4 MeV $\gamma$) & 7.&3 & $\pm$ & 0.&8\\
\multirow{2}{*}{4 $\bigg{\lbrace}$} 
& $^{13}{\rm C}(\alpha,{\it n})^{16}{\rm O}$, 1st e.s. (6.05 MeV $e^{+}e^{-}$) & 15.&9 & $\pm$ & 3.&3\\
& $^{13}{\rm C}(\alpha,{\it n})^{16}{\rm O}$, 2nd e.s. (6.13 MeV $\gamma$) & 3.&7 & $\pm$ & 0.&7\\
5 & Fast neutron and atmospheric neutrino & \multicolumn{5}{c}{$<$ 12.3}\\
%\hline
Total  \hspace{-0.3cm} &  & 325.&9 & $\pm$ & 26.&1\\
\hline
\hline
\end{tabular}
\end{table}
\end{center}

\begin{figure}
\vspace{-0.3cm}
\begin{center}
\includegraphics[angle=270,width=1.0\columnwidth]{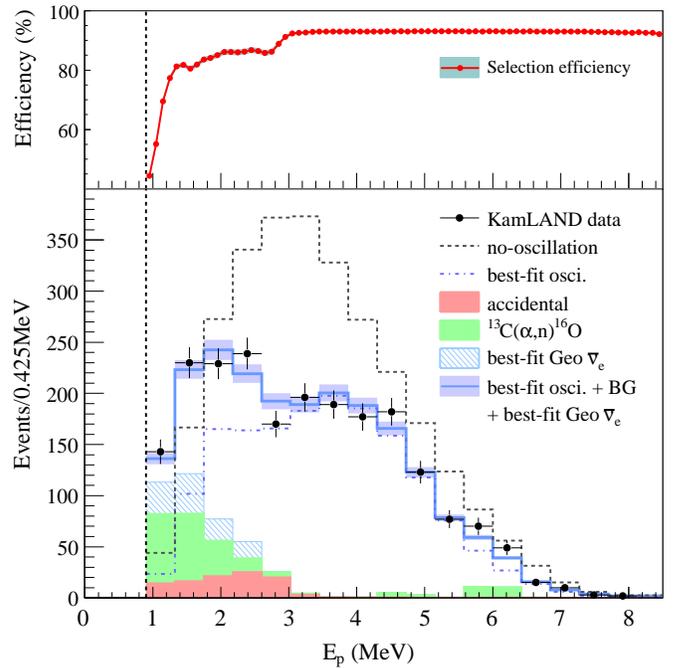}
\end{center}
\vspace{-0.2cm}
\caption[]{Prompt energy spectrum of $\overline{\nu}_{e}$ candidate events above 0.9 MeV energy threshold (vertical dashed line).  The data together with the background and reactor $\overline{\nu}_{e}$ contributions fitted from an unbinned maximum-likelihood three-flavor oscillation analysis are shown in the main panel. The number of geo-$\overline{\nu}_{e}$'s is unconstrained in the fit. The shaded background histograms are cumulative. The top panel shows the energy-dependent selection efficiency; each point is the weighted average over the five time periods described in the text.}
\label{figure:spectrum}
\end{figure}
\vspace{-1cm}
\begin{figure*}
\begin{center}
\includegraphics[angle=270,width=0.98\columnwidth]{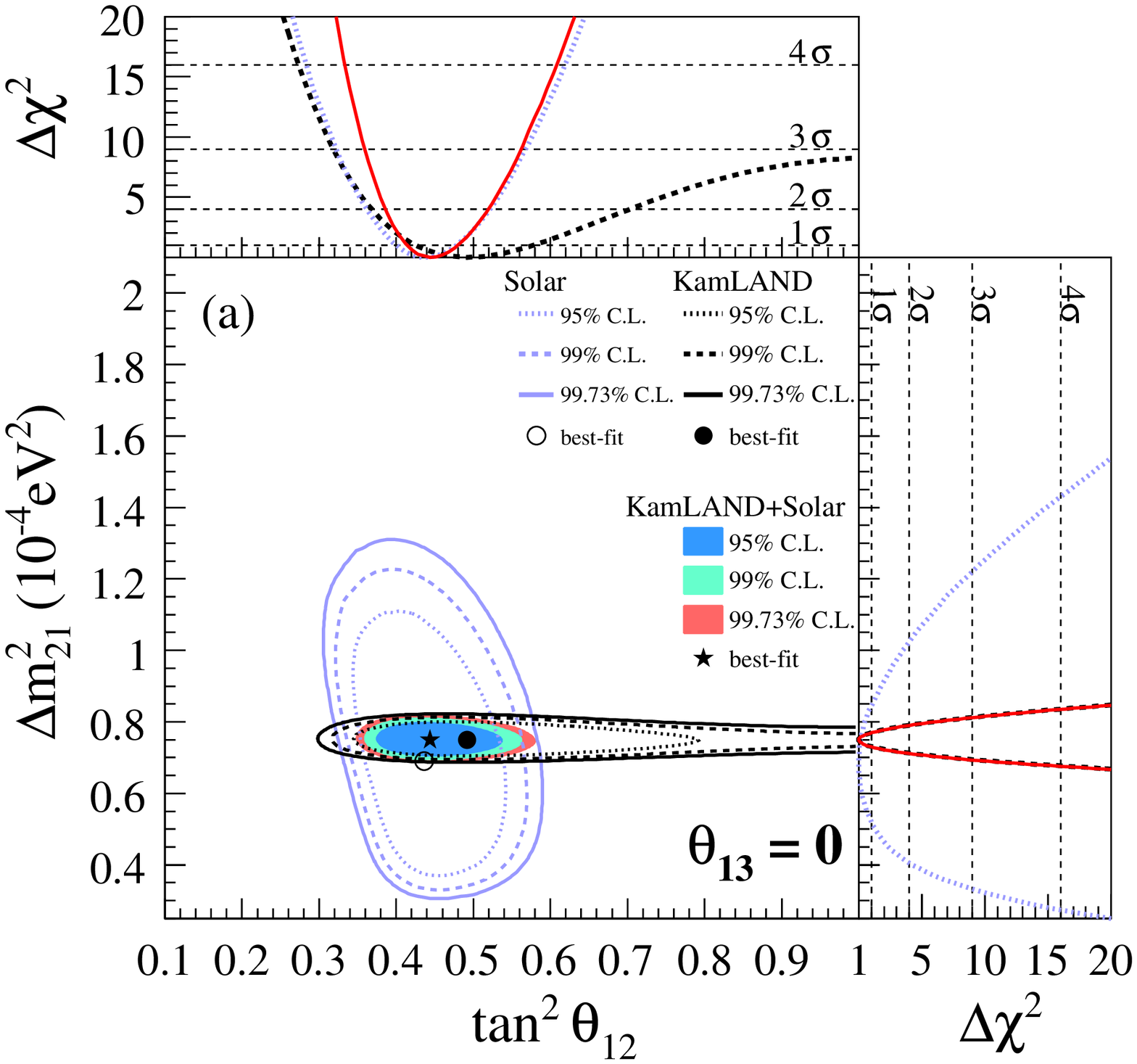}
\hspace{6mm}
\includegraphics[angle=270,width=0.98\columnwidth]{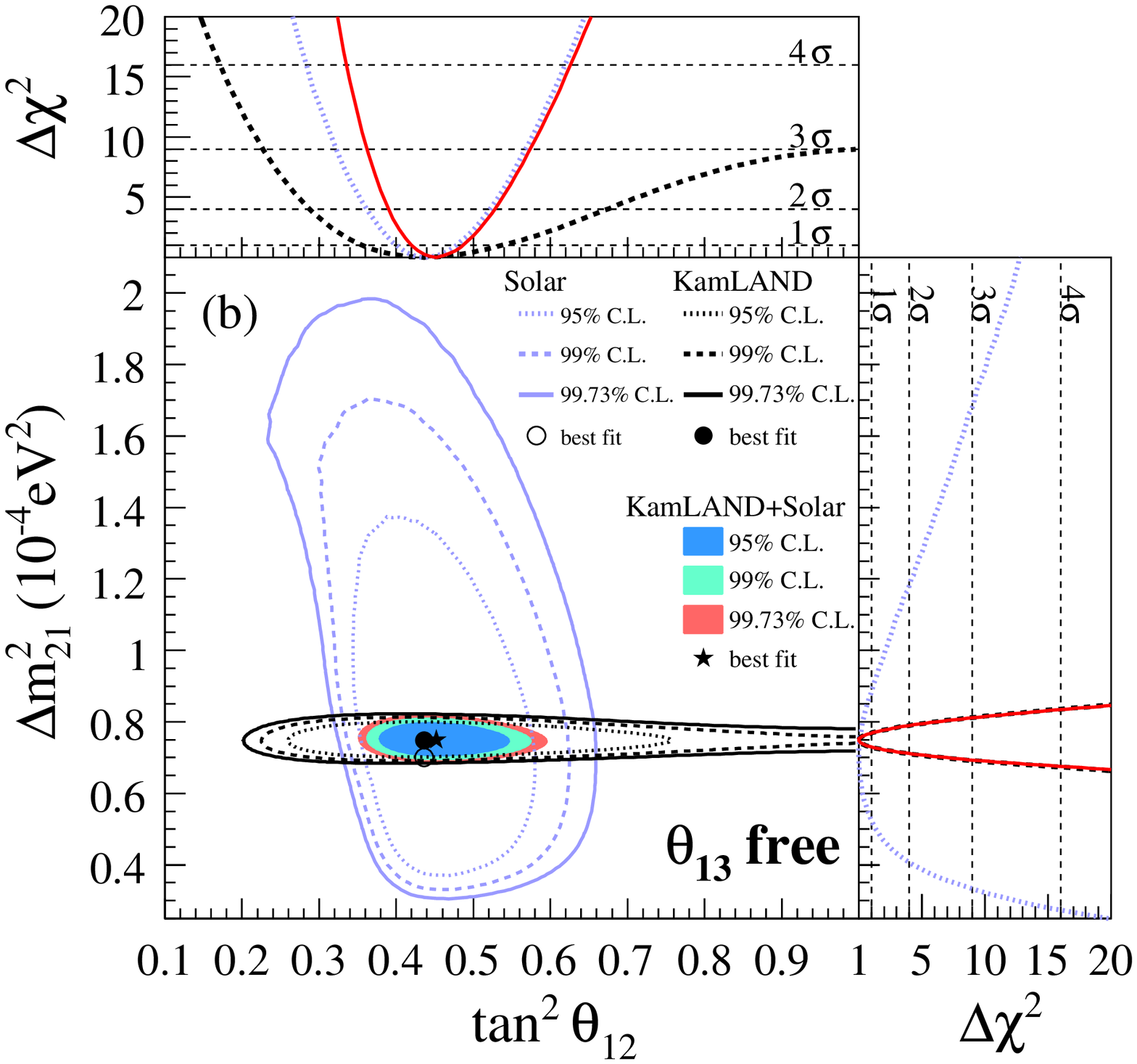}
\end{center}
\vspace{-0.2cm}
\caption[]{Allowed regions projected in the ($\tan^{2} \theta_{12}$, $\Delta m^{2}_{21}$) plane, for solar and KamLAND data from (a) the two-flavor oscillation analysis ($\theta_{13}=0$) and (b) the three-flavor oscillation analysis, where $\theta_{13}$ is a free parameter. The shaded regions are from the combined analysis of the solar and KamLAND data. The side panels show the $\Delta \chi^{2}$ profiles projected onto the $\tan^{2} \theta_{12}$ and $\Delta m^{2}_{21}$ axes.}
\label{figure:contour2}
\end{figure*}

  Antineutrino DC-pair candidates are selected by performing the following series of first-level cuts: (i) prompt energy: $0.9 < E_{\rm p} ({\rm MeV}) < 8.5 ~ $; (ii) delayed energy: $1.8 ~ < E_{\rm d}({\rm MeV}) < 2.6$ (capture on $p$), or $4.4 ~ < E_{\rm d}({\rm MeV}) < 5.6 ~$ (capture on $^{12}{\rm C}$); (iii) spatial correlation of prompt and delayed events: $\Delta R ({\rm m}) < 2.0 ~ $; (iv) time separation between prompt and delayed events: $0.5 ~  < \Delta T (\mu{\rm s})<  1000 ~ $; and (v) fiducial volume radii: $R_{\rm p}, R_{\rm d} ({\rm m}) < 6.0 ~$.  

In order to increase the ratio of signal to accidental-background, a second-level cut is performed using a likelihood discriminator, $L = \frac{f_{\overline{\nu}_{e}}}{f_{\overline{\nu}_{e}} + f_{acc}}$.  Here $f_{\overline{\nu}_{e}}$ and $f_{acc}$  are the probability density functions (PDFs) for $\overline{\nu}_{e}$ DC pairs and accidental DC pairs, respectively; both PDFs are functions of the 6 DC-pair parameters: $E_{\rm p}$, $E_{\rm d}$, $\Delta R$, $\Delta T$, $R_{\rm p}$, $R_{\rm d}$. The PDF for accidental DC pairs can be evaluated directly from the data with an off-time cut; we use  $10 ~ {\rm ms}  < \Delta T <  20 ~{\rm s}$.   To utilize the variation in the accidental DC rate with time, the full data set is divided into five periods and the corresponding $f_{acc}$ is computed for each.  The PDF for $\overline{\nu}_{e}$ DC pairs is calculated with a Monte Carlo (MC) simulation.  The systematic error in the simulated PDF is evaluated by comparing simulated calibration data to real calibration data for the $^{68}$Ge and $^{241}$Am$^{9}$Be sources. 

For each 0.1 MeV interval in prompt energy, we choose $L_{cut}(E_{\rm p})$ to maximize $\frac{S}{\sqrt{S + B_{acc}}}$, where $S$ and $B_{acc}$ are the expected number of $\overline{\nu}_{e}$ and accidental DC pairs, respectively, with $L(E_{\rm p}) > L_{cut}(E_{\rm p})$. To exploit the time variation of both the signal and background, the optimal $L_{cut}(E_{\rm p})$ is determined for each of the five time periods. Finally, only DC pairs with $L(E_{\rm p}) > L_{cut}(E_{\rm p})$ are selected.  The efficiency and uncertainty of the cut are evaluated for each period using the MC; the \mbox{$E_{\rm p}$-dependent} efficiency, averaged over the five time periods, is shown in the top panel of Fig. \ref{figure:spectrum}. A no-oscillation input spectrum is used to generate $f_{\overline{\nu}_{e}}$.  The effect of using an oscillated $\overline{\nu}_{e}$ spectrum was checked with various trial values of ($\theta_{12},  \Delta m^{2}_{21}$) and found not to greatly affect the selection.  The number of accidental DC pairs remaining after all cuts is determined to be $102.5 \pm 0.1$. The dominant contributors to these accidental DC pairs are 2.6 MeV $\gamma$-rays from external $^{208}$Tl $\beta$-decays. 

In addition to accidental background events, there are other processes which produce background DC pairs.  The $^{13}{\rm C}(\alpha,{\it n})^{16}{\rm O}$ nuclear reaction in the LS is the largest such background.  The dominant $\alpha$ source is $^{210}$Po, a long-lived daughter nucleus of $^{222}$Rn. This reaction produces neutrons with energies up to 7.3 MeV, and mostly contributes DC pairs with prompt energies below 2.6 MeV.  By counting the quenched scintillation signals from the 5.3 MeV $\alpha$ particles, we find \mbox{$(5.95 \pm 0.29) \times 10^{9}$} $\alpha$-decays in full data set.
The rate of the $^{13}{\rm C}(\alpha,{\it n})^{16}{\rm O}$ background and its prompt energy spectrum is estimated by simulation.  The total cross section and final-state partial cross sections for $^{16}{\rm O}$, $\sigma_{i}$ (where $i =0\,,1\,,2$ for the ground, first and second excited states of $^{16}{\rm O}$), are based on~\cite{Harissopulos2005, JENDL2005}, but the relative normalizations of the $\sigma_{i}$ were tuned by an {\it in-situ} calibration using a \mbox{$^{210}$Po$^{13}$C} source~\cite{McKee2008}. 
The data require $\sigma_0$ and $\sigma_{1}$ be scaled by 1.05 and 0.6, respectively, while no scaling is required for $\sigma_{2}$.  Including the uncertainty on the number of $\alpha$-decays, we assign an uncertainty of 11\% for the ground state and 20\% for the excited states. We estimate that the total number of $^{13}{\rm C}(\alpha,{\it n})^{16}{\rm O}$ DC pairs remaining in the full data set after the first- and second-level cuts is $198.6 \pm 23.0$. DS-2, which benefited from reduced $^{210}\rm{Po}$ contamination due to LS purification, contributes only 7\% of the $^{13}{\rm C}(\alpha,{\it n})^{16}{\rm O}$ events after all selection cuts.

Delayed-neutron beta emitters $^{9}$Li and $^{8}$He, which are produced in the LS by cosmic-ray muons, also generate DC pairs~\cite{Abe2010}.  They are removed by a 2-s veto of the entire fiducial volume after LS showering muons, which generate more than $10^{6}$ photoelectrons in the LS, and poorly reconstructed LS muons.  In the case of nonshowering, well-reconstructed LS muons, the 2-s veto is applied only within a 3-m-radius cylinder around the muon track in order to minimize the exposure loss from the veto.  From a fit to the time delay between prompt DC events and their preceding LS muons, we estimate the background remaining after the veto and DC selection cuts is $24.8 \pm 1.6$ events. 

Fast neutrons and atmospheric neutrinos are also a possible source of DC pairs. Fast neutrons generated in the material outside the OD may scatter into the ID, and subsequent coincidence signals in the LS from prompt neutron scatter and delayed capture sometimes pass the $\overline{\nu}_{e}$ DC signal selection criteria.   Monte Carlo studies of neutron generation outside the ID~\cite{Abe2010} indicate that fast neutrons are generated primarily by cosmic-ray muons.   A 2-ms veto after OD-tagged muons mostly eliminates fast neutron DC pairs.  The residual background due to the OD tagging inefficiency and muons that pass nearby but do not enter the OD is estimated from simulation.   Atmospheric neutrino backgrounds are evaluated using the  NUANCE software~\cite{Casper2002} to simulate neutrino interactions and related processes.  Both atmospheric neutrino and fast neutron DC pairs are assumed to have a flat prompt energy spectrum in the energy range of the present analysis, and are estimated to contribute less than 12.3 candidates in total after all selection cuts.

Geo-$\overline{\nu}_{e}$ fluxes at Kamioka can be calculated based on a reference Earth model~\cite{Enomoto2007} which assumes a radiogenic heat production rate of 16 TW from the decay chains of U and Th.  Including neutrino oscillation effects, this model predicts 85 and 21 events in the full data set from U and Th decays, respectively.  However, since the estimate of the geo-$\overline{\nu}_{e}$ yield is highly dependent on the Earth model, the event rates from the U and Th decay chains are not constrained in the oscillation analysis; only the prompt energy spectrum shapes, which are independent of the Earth model, are used to constrain their contribution. 
A possible background contribution from a hypothetical reactor-$\overline{\nu}_{e}$ source at the Earth's center, motivated by \cite{Herndon2003} and investigated in \cite{Bellini2010}, is neglected in this analysis. 

After all selection cuts, we expect, in the absence of $\overline{\nu}_{e}$ disappearance, $2879 \pm 118$ events from reactor $\overline{\nu}_{e}$, and $325.9 \pm 26.1$ events from the backgrounds, as summarized in Table~\ref{table:background}.  The observed number is 2106 events. 

\section{Oscillation analysis}
\label{section:Analysis}
The KamLAND data is analyzed based on an unbinned maximum-likelihood method.  The $\chi^{2}$ is defined by 
\begin{eqnarray}
\label{equation:chi2}
\chi^{2} & = & \chi^{2}_{\rm rate}(\theta_{12}, \theta_{13}, \Delta m^{2}_{21}, N_{\rm BG1\rightarrow5}, N^{\rm{geo}}_{{\rm U},{\rm Th}}, \alpha_{\rm 1\rightarrow4}) \nonumber \\
& & - 2 \ln L_{\rm shape} (\theta_{12}, \theta_{13}, \Delta m^{2}_{21}, N_{\rm BG1\rightarrow5}, N^{\rm{geo}}_{{\rm U},{\rm Th}}, \alpha_{\rm 1\rightarrow4}) \nonumber \\
& & + \chi^{2}_{\rm BG}(N_{\rm BG1\rightarrow5}) + \chi^{2}_{\rm syst}(\alpha_{\rm 1\rightarrow4})\;.
\end{eqnarray}
The terms are, in order: the $\chi^{2}$ contribution for (i) the total rate, (ii) the prompt energy spectrum shape, (iii) a penalty term for backgrounds, and (iv) a penalty term for systematic uncertainties. $N_{\rm BG1\rightarrow5}$ are the expected background levels discussed in Sec.~\ref{section:cutsAndBkg}, and $N^{\rm{geo}}_{{\rm U},{\rm Th}}$ are the contributions expected from U and Th geo-$\overline{\nu}_{e}$'s. $N_{\rm BG1\rightarrow5}$ are allowed to vary in the fit but are constrained with the penalty term (iii) using the estimates summarized in Table~\ref{table:background}.  $N^{\rm{geo}}_{{\rm U},{\rm Th}}$ are free parameters and are unconstrained to avoid any Earth model dependence.  The $\alpha_{\rm 1\rightarrow4}$ parametrize the uncertainties on the reactor $\overline{\nu}_{e}$ spectra and energy scale, the event rate, and the energy dependent efficiencies; these parameters are allowed to vary in the analysis but are constrained by term (iv).  The background energy scale uncertainties are estimated to contribute at most an additional 0.5\% to the error on the event rate and are neglected in this analysis.  The prompt energy spectrum shape likelihood term (ii) is evaluated as a function of the candidate event time.  The detailed knowledge of the time evolution of the total reactor $\overline{\nu}_{e}$ spectrum and effective baseline, afforded by the reactor fuel composition and power data provided by the Japanese reactor operators, is thus fully utilized in the analysis.  Variations in the total observed spectrum shape with time due to changes in the background levels---especially the $^{13}{\rm C}(\alpha,{\it n})^{16}{\rm O}$ reduction from the LS purification---are also exploited by this term.  The spectrum shape likelihood term allows an Earth-model-independent constraint of the geo-$\overline{\nu}_{e}$ contribution since the U and Th decay spectra are known independently of the Earth model. A $\chi^{2}$ scan of the ($\theta_{12}, \theta_{13}, \Delta m^{2}_{21}$) oscillation parameter space is carried out, minimizing $\chi^{2}$ with respect to $N_{\rm BG1\rightarrow5}$, $N^{\rm{geo}}_{{\rm U},{\rm Th}}$, and $\alpha_{\rm 1\rightarrow4}$. 
\begin{figure}[t]
\begin{center}
\includegraphics[angle=270,width=0.98\columnwidth]{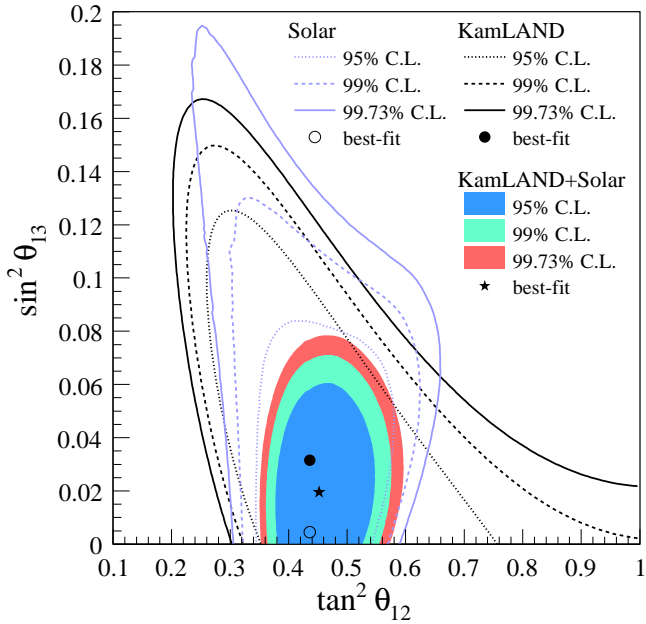}
\end{center}
\vspace{-0.2cm}
\caption[]{Allowed regions from the solar and KamLAND data projected in the ($\tan^{2} \theta_{12}$, $\sin^{2} \theta_{13}$) plane for the three-flavor analysis. }
\label{figure:contour3}
\end{figure}
\begin{figure}[t]
\vspace{-0.41cm}
\begin{center}
\includegraphics[angle=270,width=1.0\columnwidth]{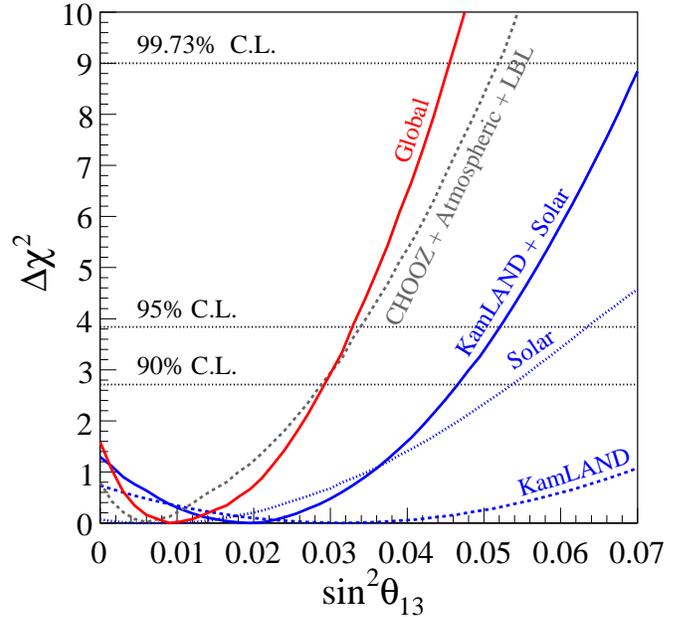}
\end{center}
\vspace{-0.25cm}
\caption[]{$\Delta \chi^{2}$-profiles projected onto the $\sin^{2} \theta_{13}$ axis for different combinations of the oscillation data floating the undisplayed parameters ($\tan^{2} \theta_{12}$, $\Delta m^{2}_{21}$).
}
\label{figure:chi2}
\end{figure}

In our analysis of the solar neutrino data, we include the rates in the chlorine~\cite{Cleveland1998} and gallium~\cite{Abdurashitov2009} experiments, Borexino~\cite{Arpesella2008b}, SNO III~\cite{Aharmim2008}, the zenith spectra in Super-Kamiokande phase~I~\cite{Hosaka2006}, and the day-night spectra in SNO phase~I and II~\cite{Aharmim2010}.  The measured fluxes are compared with the high-metallicity standard solar model predictions (GS98)~\cite{Grevesse1998}.

For the three-flavor KamLAND-only analysis, without any constraints on $\theta_{13}$ from other oscillation experiments, the best-fit oscillation parameter values are  $\Delta m^{2}_{21} = 7.49^{+0.20}_{-0.20} \times 10^{-5} ~ {\rm eV}^{2}$, $\tan^{2} \theta_{12} = 0.436^{+0.102}_{-0.081}$ and \mbox{$\sin^{2} \theta_{13} = 0.032^{+0.037}_{-0.037}$ ($< 0.094$ at the 90\% C.L.)}. The two-flavor oscillation treatment using \mbox{Eq.~(\ref{equation:chi2})}, as presented previously in ~\cite{Abe2008}, is a special case of the three-flavor treatment with $\theta_{13} = 0$. For this case the best-fit oscillation parameters from the KamLAND-only analysis are $\Delta m^{2}_{21} = 7.50^{+0.20}_{-0.20} \times 10^{-5} ~ {\rm eV}^{2}$ and $\tan^{2} \theta_{12} = 0.492^{+0.086}_{-0.067}$. In the KamLAND data, $\theta_{13}$ is expected to contribute only an energy-independent event rate suppression and we find almost no effect on the $\Delta m^{2}_{21}$ measurement when $\theta_{13}$ is included as a free parameter.  Figure~\ref{figure:spectrum} shows the prompt energy spectrum of candidate events in KamLAND together with the best-fit background and reactor $\overline{\nu}_{e}$ spectra for the three-flavor fit to the KamLAND data.  The fit estimates 82 and 26 events from U and Th geo-$\overline{\nu}_{e}$'s, respectively, in agreement with the reference model. 

Figure~\ref{figure:contour2} compares the allowed regions in the \mbox{$(\tan^{2}\theta_{12}, \Delta m^{2}_{21})$} plane from the two- and three-flavor oscillation analyses.  
We find [Fig. ~\ref{figure:contour2}(a)] that the allowed region from the solar data is in agreement with the KamLAND data, and the small tension between the two-flavor best-fit values of $\theta_{12}$, discussed previously in~\cite{Fogli2008, Garcia2010}, has eased.  Assuming {\it CPT} invariance, the two-neutrino oscillation parameter values from a combined analysis of the solar and KamLAND data are $\tan^{2} \theta_{12} = 0.444^{+0.036}_{-0.030}$ and $\Delta m^{2}_{21} = 7.50^{+0.19}_{-0.20} \times 10^{-5} ~ {\rm eV}^{2}$.  For the three-flavor analysis combining the solar and KamLAND data, the best-fit parameter values are $\tan^{2} \theta_{12} = 0.452^{+0.035}_{-0.033}$ and $\sin^{2} \theta_{13} = 0.020^{+0.016}_{-0.016}$; the best-fit value for $\Delta m^2_{21}$ is the same as for the two-flavor result. The best-fit values for the different data combinations and analysis approaches are summarized in Table~\ref{table:fit-values} in Appendix~A.

Figure~\ref{figure:contour3} shows the regions in the \mbox{($\tan^{2} \theta_{12}$, $\sin^{2} \theta_{13}$)} plane allowed by $\chi^{2}$ minimization with respect to $\Delta m^{2}_{21}$ for each analysis.  The reduction of the best-fit value of $\tan^{2}\theta_{12}$ for the three-flavor KamLAND-only analysis relative to the two-flavor KamLAND analysis (Fig. ~\ref{figure:contour2}) follows the anticorrelation apparent in the KamLAND contours (Fig. ~\ref{figure:contour3}). The correlation between $\theta_{12}$ and $\theta_{13}$ in the solar data is slight and the difference between the best-fit values of $\theta_{12}$ from the two-flavor and three-flavor analyses of the solar-only data is small.  

 \begin{figure}
 \begin{minipage}{1.0\columnwidth}
 \vspace{0.5cm}
 \begin{center}
 \includegraphics[angle=270,width=1.0\columnwidth]{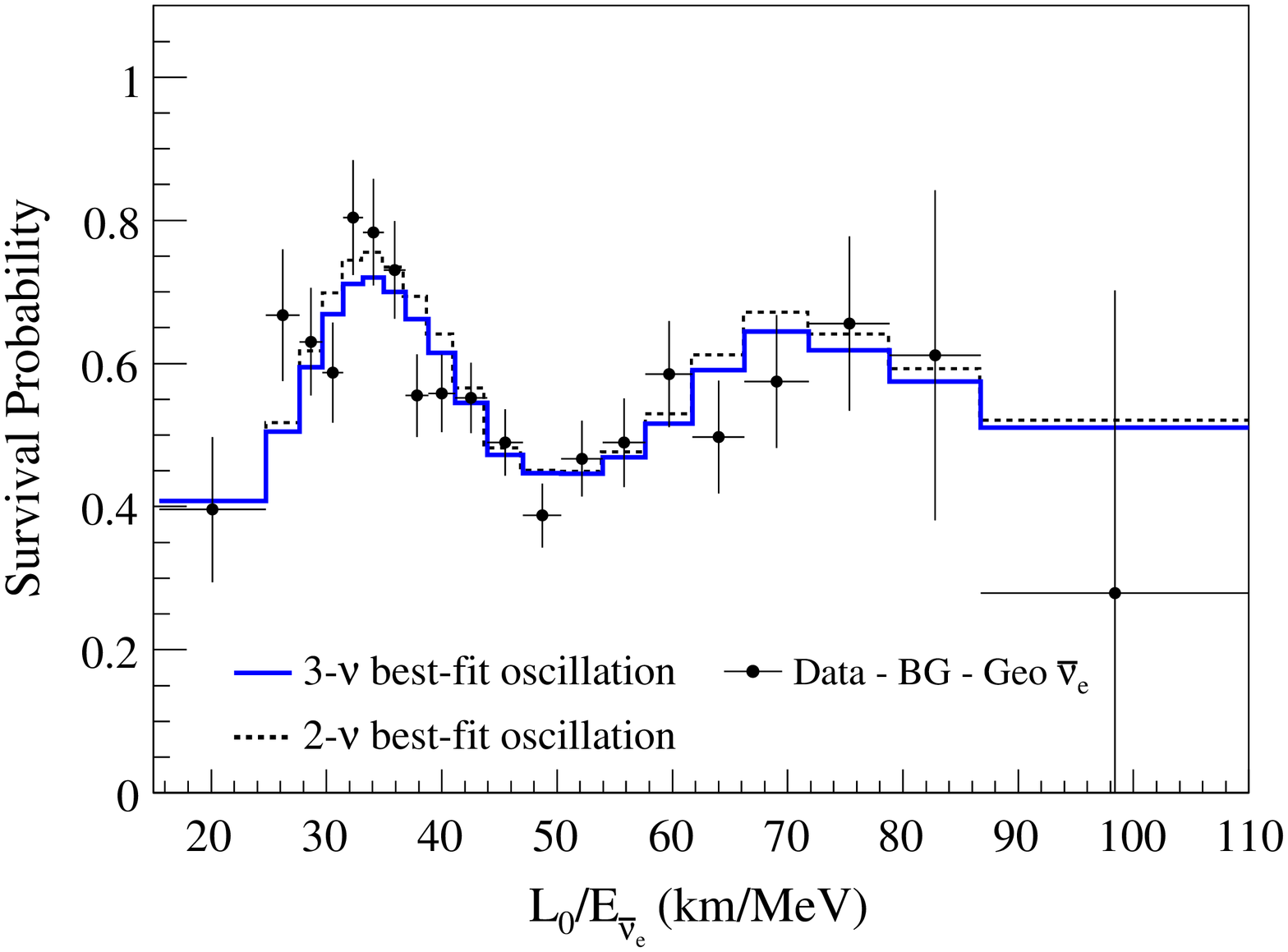}
 \end{center}
 \caption[]{Ratio of the observed $\overline{\nu}_{e}$ spectrum to the expectation for no-oscillation versus $L_{0}/E$ for the KamLAND data. $L_{0} = 180~{\rm km}$ is the flux-weighted average reactor baseline.  The 2-$\nu$ and 3-$\nu$ histograms are the expected distributions based on the best-fit parameter values from the two- and three-flavor unbinned maximum-likelihood analyses of the KamLAND data.}
 \label{figure:LE}
 \end{minipage}
 \begin{minipage}{1.0\columnwidth}
 \vspace{1.0cm}
 \begin{center}
 \includegraphics[angle=270,width=1.0\columnwidth]{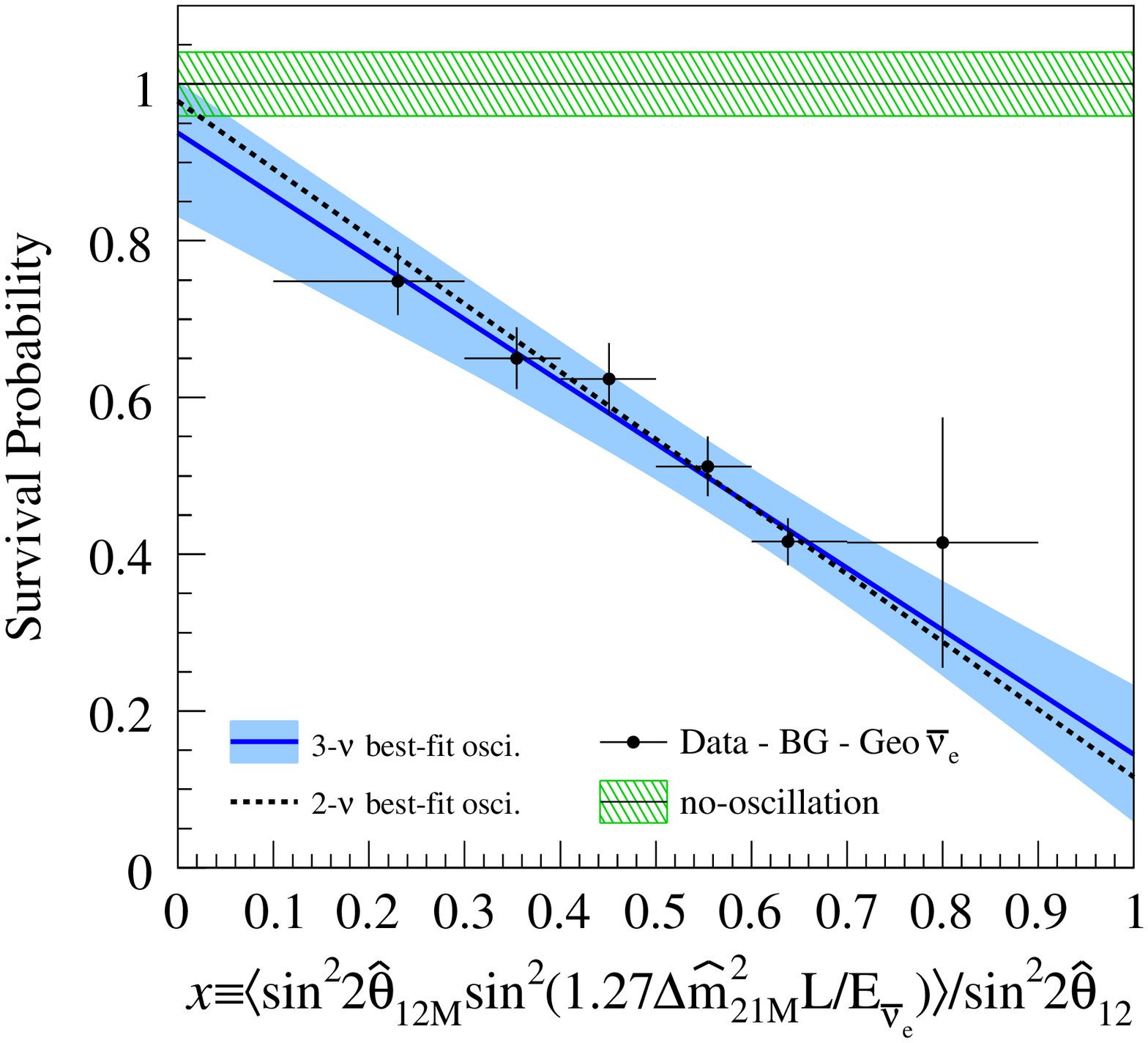}
 \end{center}
 \vspace{-0.2cm}
 \caption[]{Survival probability of reactor $\overline{\nu}_{e}$ versus
\mbox{$x  \equiv \langle \sin^{2}2\hat{\theta}_{12M}\sin^2(1.27 \Delta \hat{m}^2_{21M}L/E)\rangle /\sin^22\hat{\theta}_{12}$}. The angle bracket indicates the weighted average over reactor baseline $(L_{i})$ and original neutrino energies $(E)$. The points are the survival probability for the KamLAND data.   The 3-$\nu$ line and $1\sigma$ C.L. region are calculated using the unbinned maximum-likelihood fit to the KamLAND data.  The 2-$\nu$ line is calculated from the two-flavor unbinned maximum-likelihood KamLAND analysis. The $1\sigma$ C.L. band for the 2-$\nu$ case is not shown but is similar in magnitude to the no-oscillation case shown at $P=1.0$.
 }
 \label{figure:P_LE}
 \end{minipage}
 \end{figure}
 Figure \ref{figure:chi2} shows $\Delta \chi^{2}$-profiles projected onto the $\sin^{2} \theta_{13}$ axis for different combinations of the data. The analysis of the KamLAND data gives $\sin^{2} \theta_{13} = 0.032^{+0.037}_{-0.037}$ ($< 0.094$ at the 90\% C.L.), and the combined analysis of the solar and KamLAND data gives $\sin^{2} \theta_{13} = 0.020^{+0.016}_{-0.016}$.
The constraint on nonzero $\theta_{13}$ from the combined KamLAND and solar analysis is comparable to the constraint from the combined analysis of CHOOZ, atmospheric, and long-baseline accelerator (LBL, i.e., K2K and MINOS) experiments presented in \cite{Garcia2010a}, which includes the recent $\nu_{e}$\ appearance result from MINOS~\cite{Adamson2010}.
In the solar + KamLAND analysis the preference for nonzero $\theta_{13}$ comes mostly from the KamLAND data.
All oscillation data favor a positive $\theta_{13}$, although the current statistical power is poor.
For a global analysis combining our updated KamLAND + solar analysis with the combined CHOOZ, atmospheric, and LBL (appearance + disappearance) analysis from \cite{Garcia2010a}, we find $\sin^{2} \theta_{13} = 0.009^{+0.013}_{-0.007}$. Our global result is very similar to the global analysis carried out by \cite{Garcia2010a} with the previous KamLAND data set but the significance of nonzero $\theta_{13}$ is reduced slightly to the 79\% C.L. 

\section{ Visualization of  the survival probability}
\label{section:vis}
Figures ~\ref{figure:LE} and  ~\ref{figure:P_LE} illustrate different aspects of the survival probability for the KamLAND data.  The data points in Fig. ~\ref{figure:LE} are the ratio of the observed reactor $\overline{\nu}_{e}$ spectrum to that expected in the case of no oscillation plotted as a function of $L_{0}/E$, where $L_{0}$ ($L_{0} = 180~\rm{km})$ is the flux-weighted average reactor baseline. The oscillatory structure arising from the $\sin^2(1.27 \Delta m^{2}_{21}L/E)$ term is clear, but is distorted because the reactor sources are distributed across multiple baselines.  We also overlay in the figure the expected oscillation curves based on the best-fit parameters from the two- and three-flavor unbinned maximum-likelihood analyses discussed previously. The suppression of the oscillation amplitude is slightly larger for the nonzero $\theta_{13}$ case. 

To focus on $\theta_{12}$ and $\theta_{13}$ effects in the data, we introduce a parameter $x(E_{\rm p},t)$ defined by
\begin{widetext}
\begin{eqnarray}
x(E_{\rm p},t) &=&\frac{1}{\sin^2{2\hat{\theta}_{12}}}\bigg{[}\frac{1}{N_{\rm{no\mathchar`-osc}}(E_{\rm p},t)}\sum_{i}^{\rm reactors} \int \mathrm{d}E ~ \sin^22\hat{\theta}_{12M}\sin^2\bigg{(}\frac{1.27\Delta \hat{m}^2_{21M}L_{i}}{E}\bigg{)}P_{\rm{R}}(E_{\rm p},t,E)\frac{S_{i}(E,t)}{4\pi L^2_{i}} \bigg{]}\, \label{equation:x-explicit}\\
&\equiv&\frac{1}{\sin^2{2\hat{\theta}_{12}}}\bigg{\langle}\sin^22\hat{\theta}_{12M}\sin^2\bigg{(}\frac{1.27\Delta \hat{m}^2_{21M}L}{E}\bigg{)}\bigg{\rangle}\,, \label{equation:x-notation}
\end{eqnarray}
\end{widetext}
where 
\begin{eqnarray}
N_{\rm{no\mathchar`-osc}}(E_{\rm p},t)=\sum_{i}^{\rm reactors}\int \mathrm{d}E ~ P_{\rm{R}}(E_{\rm p},t,E)\frac{S_{i}(E,t)}{4\pi L^2_{i}} \label{equation:defineSDet}
\end{eqnarray}
is the number of candidates with prompt energy $E_{\rm p}$ expected in the absence of neutrino oscillation from all reactors at time $t$ at KamLAND; the index $i$ labels the reactor source; $L_{i}$ and $S_{i}(E,t)$ are, respectively, the baseline and the neutrino spectrum at time $t$ of reactor $i$; and $P_{\rm{R}}(E_{\rm p},t,E)$ is the probability that a $\overline{\nu}_{e}$ with energy $E$ will be detected at KamLAND with prompt energy $E_{\rm p}$. $P_{\rm{R}}$ includes the number of target protons, the inverse beta-decay cross section, and the time-dependent detector response function. ($\hat{\theta}_{12},\Delta \hat{m}^2_{21}$) are the best-fit values from the two-flavor unbinned analysis, and  ($\hat{\theta}_{12M}$,$\Delta \hat{m}^2_{21M}$) are the matter-modified oscillation parameters calculated with those best-fit values.  The angle bracket notation in \mbox{Eq.~(\ref{equation:x-notation})} indicates the weighted average over reactor baselines $L_{i}$ and neutrino emission energies $E$, written explicitly in \mbox{Eq.~(\ref{equation:x-explicit})}.
For the region of ($\Delta m^2_{21},\theta_{12},\theta_{13}$) parameter space close to $\Delta \hat{m}^2_{21}$, all the information about the reactors, detector-related effects, and matter modification is contained in the parameter $x$.  With this definition, the survival probability may be written as a linear function of $x$, $P(E_{\rm p},t)=A - B\cdot x(E_{\rm p},t)\,,$  
where $A = (\cos^{4}\theta_{13} + \sin^{4}\theta_{13}) $ and $B = \cos^{4}\theta_{13}\sin^{2}2\theta_{12}$\,. $\theta_{13}$ effects are predominantly encoded in $A$, whereas $\theta_{12}$ effects dominate the slope $B$. This linear relationship is illustrated in Fig. \ref{figure:P_LE}. The points there are the survival probability for KamLAND events binned as a function of $x$. Also shown are lines where $A$ and $B$ have been calculated using the best-fit values from the two- and three-flavor unbinned maximum-likelihood analyses of the KamLAND data.  The axis intercept at $x=0$ of the best-fit 3-$\nu$ line is less than one, illustrating the slight indication of positive $\theta_{13}$ from the unbinned likelihood analysis.  Any further improvement in the significance of the $\theta_{13}$ investigation with KamLAND requires reduced systematic uncertainties on the reactor flux and increased detector exposure.  A binned analysis based on the data points in Fig. \ref{figure:P_LE} is outlined in Appendix~B.

\section{Conclusion}
\label{section:Conclusion}

An updated KamLAND reactor $\overline{\nu}_{e}$ data set was presented. The data set benefits from increased exposure and an improved background environment due to a radiopurity upgrade of the LS.  The analysis slightly hints at a nonzero $\theta_{13}$ with the available oscillation data.   In a two-flavor analysis ($\theta_{13} = 0$) of the solar and KamLAND data, the best-fit values for the oscillation parameters are $\tan^{2} \theta_{12} = 0.444^{+0.036}_{-0.030}$ and $\Delta m^{2}_{21} = 7.50^{+0.19}_{-0.20} \times 10^{-5} ~ {\rm eV}^{2}$.  In the three-flavor analysis, floating the value of $\theta_{13}$ without any constraints from the other oscillation experiments gives the solar + KamLAND best-fit values $\tan^{2} \theta_{12} = 0.452^{+0.035}_{-0.033}$, $\Delta m^{2}_{21} = 7.50^{+0.19}_{-0.20} \times 10^{-5} ~ {\rm eV}^{2}$, and $\sin^{2} \theta_{13} = 0.020^{+0.016}_{-0.016}$.  The limits on $\Delta m^{2}_{21}$ are the same for the two- and three-flavor analyses. 
All three oscillation parameters derived from the KamLAND-only antineutrino data are in good agreement with those from the solar-only neutrino data and reveal no inconsistency with {\it CPT} invariance, which was assumed for the joint fits. 
The upper limit we obtain on $\sin^{2} \theta_{13}$ is compatible with other recent work combining CHOOZ, atmospheric, and accelerator experiments. 
More definitive information on the value of $\theta_{13}$ should come from upcoming accelerator and reactor experiments.

\section*{ACKNOWLEDGMENTS}

The KamLAND experiment is supported by the Grant-in-Aid for 
Specially Promoted Research under grant 16002002
of the Japanese Ministry of Education, Culture, Sports, Science and Technology;
the World Premier International Research Center Initiative (WPI Initiative), MEXT, Japan;
and under the U.S.\ Department of Energy (DOE) Grants DE-FG03-00ER41138, DE-AC02-05CH11231, and DE-FG02-01ER41166,
as well as other DOE grants to individual institutions.
The reactor data are provided by courtesy of the following electric associations in Japan: Hokkaido, Tohoku, Tokyo, Hokuriku, Chubu, Kansai, Chugoku, Shikoku, and Kyushu Electric Power Companies, Japan Atomic Power Company, and Japan Atomic Energy Agency. 
The Kamioka Mining and Smelting Company has provided service for activities in the mine.

\vspace{1.0cm}
\section*{APPENDIX A}
The best-fit values for the different data combinations and analysis approaches are summarized in Table~\ref{table:fit-values}.
\hspace{0.5cm}
\begin{center}
\begin{table}[h]
\caption{\label{table:fit-values}
Summary of the best-fit values for $\tan^{2} \theta_{12}$ and $\sin^{2} \theta_{13}$ from two- and three-flavor neutrino oscillation analyses of various combinations of experimental data.  
}
\centering
\begin{tabular}{@{}*{4}{lccc}}
\hline
\hline
\hspace{0.3cm} Data set & Analysis method & \hspace{0.3cm} $\tan^{2} \theta_{12}$ \hspace{0.3cm} & \hspace{0.3cm} $\sin^{2} \theta_{13}$ \hspace{0.3cm} \\
\hline
KamLAND & two-flavor & $0.492^{+0.086}_{-0.067}$ & $\equiv 0$ \\
KamLAND + solar & two-flavor & $0.444^{+0.036}_{-0.030}$ & $\equiv 0$ \\
KamLAND & three-flavor & $0.436^{+0.102}_{-0.081}$ & $0.032^{+0.037}_{-0.037}$ \\
KamLAND + solar & three-flavor & $0.452^{+0.035}_{-0.033}$ & $0.020^{+0.016}_{-0.016}$ \\
Global & three-flavor & $0.444^{+0.039}_{-0.027}$ & $0.009^{+0.013}_{-0.007}$ \\
\hline
\hline
\end{tabular}
\end{table}
\vspace{-1.0cm}
\end{center}
\hspace{3.0cm}

\section*{APPENDIX B}
We consider the unbinned maximum-likelihood method presented in Sec. \ref{section:Analysis} to be the optimal approach to analyzing the KamLAND data because it takes full advantage of all the spectral and time information available.   In this appendix we outline a binned oscillation analysis which we find reproduces very well the $\Delta \chi^{2}$ contours in the ($\theta_{12},\theta_{13}$) subspace for the unbinned likelihood KamLAND-only analysis shown in \mbox{Fig. \ref{figure:contour3}}.  The binning parameter is the parameter $x$ introduced in Sec. \ref{section:vis} and defined in \mbox{Eq.~(\ref{equation:x-explicit})}. Table~\ref{table:P_LE} lists the binned data. The binned $\chi^{2}$ is defined as 
 \begin{eqnarray}
 \chi^{2} & = & \sum_{i} \left\{ \frac{p_{i} - \rho_{i}(1 + \delta_{\rm corr})}{ \sigma_{p_{i}}} \right\}^{2} + \left( \frac{\delta_{\rm corr}}{\sigma_{\rm corr}} \right)^{2}\,,
 \end{eqnarray}
\vspace{-0.5cm}
 where
 \begin{eqnarray}
 \rho_{i} & = & \cos^{4}\theta_{13}(1 - \sin^{2}2\theta_{12} \cdot \overline{x} _{i}) + \sin^{4}\theta_{13} \label{equation:defineRho}
 \end{eqnarray}
and $ \overline{x}_{i}$ is the weighted average of $x$ over bin $i$.  The pairs \mbox{($p_{i}$, $\sigma_{p_{i}}$)} are the observed survival probability, defined as the ratio of the observed events to the expectation for no oscillation,  and its uncertainty for each bin $i$, and $\delta_{\rm corr}$ is a factor needed to account for the systematic uncertainty  \mbox{($\sigma_{\rm corr} = 0.041$)} on the flux prediction. In \mbox{Eq.~(\ref{equation:defineRho})}, the vacuum $\theta_{12}$ should be used because matter corrections to $\theta_{12}$ and $\Delta m_{21}^{2}$ are included in the calculation of $x$, as shown in \mbox{Eq.~(\ref{equation:x-explicit})}. For a global analysis, the small dependence on $\Delta m_{21}^{2}$ can be ignored and the binned $\chi^{2}$ may be used for a scan over the  $(\theta_{12}, \theta_{13})$ oscillation parameter space. Comparing the $\Delta \chi^{2}$ map built using this method and that from the full unbinned analysis shown in \mbox{Fig. \ref{figure:contour3}}, the only significant deviations appear far from the best-fit point at high values of $\theta_{12}$ where constraints from the solar neutrino experiments dominate.
 
 \begin{center}
 \begin{table}[h]
 \vspace{-1.0cm}
 \caption{\label{table:P_LE} Survival probability for each bin in $x$ [defined in 
\mbox{Eq.~(\ref{equation:x-explicit})}]. 
The first column indicates the bin range of \mbox{$x \equiv \langle \sin^{2}2\hat{\theta}_{12M}\sin^{2}(1.27 \Delta \hat{m}_{21M}^{2} L / E) \rangle /\sin^{2}2{\hat{\theta}_{12}}$}. The weighted average $\overline{x}$ is given in the second column. The observed survival probability is shown in the third column. 
The uncertainties include only the statistical and background estimation uncertainties, which are assumed to be uncorrelated. In addition, the systematic uncertainty \mbox{($\sigma_{\rm corr} = 4.1\%$)} on the flux prediction needs to be included for each bin as a fully correlated uncertainty.}
\centering
\begin{tabular}{@{}*{7}{c}}
\hline
\hline
\hspace{0.5cm} $x$ range \hspace{0.5cm} & \hspace{0.3cm} $\overline{x}$ \hspace{0.3cm} &  \hspace{0.5cm} Survival probability $(p \pm \sigma_{p})$ \hspace{0.5cm} & \\
\hline
0.1-0.3 &0.230 & $0.749 \pm 0.044$\\
0.3-0.4 &0.354 & $0.650 \pm 0.039$\\
0.4-0.5 &0.451 & $0.624 \pm 0.046$\\
0.5-0.6 &0.555 & $0.512 \pm 0.038$\\
0.6-0.7 &0.638 & $0.416 \pm 0.030$\\
0.7-0.9 &0.800 & $0.415 \pm 0.160$\\
\hline
\hline
\end{tabular}
\end{table}
\end{center}

 \bibliography{ReactorNeutrino}

\end{document}